\documentclass[aps,prb,preprint,superscriptaddress]{revtex4-1}
\usepackage{graphicx,amsmath,amsfonts}
\begin{document}
\title{Orbital magnetization of insulating perovskite transition-metal oxides with the net ferromagnetic moment in the ground state}
\author{S. A. Nikolaev}
\affiliation{Department of Theoretical Physics and Applied Mathematics, Ural Federal University,
Mira str. 19, 620002 Ekaterinburg, Russia}
\author{I. V. Solovyev}
\email{SOLOVYEV.Igor@nims.go.jp}
\affiliation{Department of Theoretical Physics and Applied Mathematics, Ural Federal University,
Mira str. 19, 620002 Ekaterinburg, Russia}
\affiliation{Computational Materials Science Unit,
National Institute for Materials Science, 1-2-1 Sengen, Tsukuba,
Ibaraki 305-0047, Japan}

\date{\today}
\begin{abstract}
Modern theory of the orbital magnetization is applied to the series of
insulating perovskite transition metal oxides (orthorhombic YTiO$_3$, LaMnO$_3$, and YVO$_3$,
as well as monoclinic YVO$_3$), carrying a net ferromagnetic (FM) moment in the ground state.
For these purposes, we use an effective Hubbard-type model,
derived from the first-principles electronic-structure calculations and
describing the behavior of magnetically active states near the Fermi level.
The solution of this model in the mean-field Hartree-Fock approximation with the relativistic spin-orbit coupling
typically gives us a distribution of the local
orbital magnetic moments, which are related to the site-diagonal part of the density matrix $\hat{\cal D}$
by the ``standard'' expression
$\boldsymbol{\mu}^0 = - \mu_{\rm B} \mathrm{Tr}  \{ \hat{\textbf{L}} \hat{\cal D}  \}$ and which are
usually well quenched by the crystal field.
In this work, we evaluate ``itinerant'' corrections $\Delta \boldsymbol{\cal M}$ to the net FM moment,
suggested by the modern theory.
We show that these corrections are small and
in most cases can be neglected. Nevertheless,
the most interesting aspect of our analysis is that, even for these compounds,
which are typically regarded as normal Mott insulators,
the ``itinerant'' corrections reveal a strong $\textbf{k}$-dependence
in the reciprocal space, following the behavior of Chern invariants.
Therefore, the small value of $\Delta \boldsymbol{\cal M}$ is the result of strong cancelation of relatively large
contributions, coming from different parts of the Brillouin zone.
We discuss details as well as possible implications of this cancelation, which depends on the
crystal structure as well as the type of the magnetic ground state.
\end{abstract}

\maketitle
\section{\label{sec:Intro} Introduction}

  Orbital magnetism is one of the oldest and most fundamental phenomena.
All our present understanding of magnetism developed from the classical concept of orbital motion,
which is much older than the concept of spin. The orbital magnetization can be
probed by
many experimental techniques, including susceptibility measurement, electron paramagnetic resonances, x-ray magnetic circular dichroism,
neutron diffraction, etc.\cite{White,Syncrotron,Lander}

  At the same time, the orbital magnetism appears to be one of the most difficult and challenging problems for the theory,
especially when it comes to the level of first-principles electronic structure calculations. If the methods of spin magnetism
are relatively well elaborated, the study of orbital magnetism is sometimes regarded to be on a primitive stage.
There are two reasons for it.

  The first one is that the spin magnetism, in principle, allows for the description starting from the limit of
homogeneous electron gas, which is widely used as an approximation for the
exchange-correlation energy (the so-called local spin density approximation or LSDA) in the spin-density functional theory (SDFT).
On the contrary, the orbital magnetism implies
some inhomogeneities of the medium, being associated with either the spin-orbit (SO) interaction or
the external vector potential, which are necessary to induce the magnetization.\cite{remark2}
Therefore, for the correct description of orbital magnetization on the level of first-principles
electronic structure calculations, it is essential to go beyond the
homogeneous electron gas limit. Furthermore, there may be even more fundamental problem, related to the fact that
the Kohn-Sham SDFT (even the exact one) does not necessary guarantee to yield correct orbital currents and, therefore,
the orbital magnetization, which is defined in terms of these currents.\cite{VignaleRasolt} This means that
the orbital magnetization (or any related to it quantity)
should be treated as an independent variational degree of freedom in the density functional theory (DFT).\cite{Jansen}
Historically, this problem in calculations of the orbital magnetization was noticed first, and
on earlier stages all the efforts were mainly concentrated on the improvement of SDFT,
by introducing different kinds of
semi-empirical orbital functionals (Refs.~\onlinecite{OPB,Norman,LDAU,Minar})
or moving in the direction of \textit{ab initio} current SDFT (Ref.~\onlinecite{Ebert}).
Most of these theories emphasized the local character of the orbital magnetization,
implying that
(i) it can be computed using the standard expression
\begin{equation}
\boldsymbol{\mu}^0 = - \mu_{\rm B} \mathrm{Tr}  \{ \hat{\textbf{L}} \hat{\cal D}  \}
\label{eqn:canonical}
\end{equation}
for the
expectation value of the angular momentum operator $\hat{\textbf{L}}$ in terms of the
site-diagonal part of the density matrix $\hat{\cal D}$, where $\mu_{\rm B} = e \hbar/2mc$ is the Bohr magneton
in terms of the electron charge ($-$$e$), its mass ($m$), the Plank constant ($\hbar$), and the
velocity of light ($c$);
and (ii) the effect of exchange-correlation interactions on $\boldsymbol{\mu}^0$
can be also treated in the local form, by considering only properly screened on-site interactions
and the same site-diagonal elements of the density matrix (Ref.~\onlinecite{OPB,Norman,LDAU}) or
of the lattice Green function (Ref.~\onlinecite{Minar}).
Even today, the problem of how to
``decorate'' DFT in order to describe properly the effects of orbital magnetism in solids
is largely unresolved and continues to be one of the most important and interesting issues.

  Nevertheless, the new turn in the theory of orbital magnetism was not directly related to fundamentals of DFT.
It was initiated by another fundamental question of how the orbital magnetization should be computed
for extended periodic systems.
This new direction, which we will refer to as
the ``modern theory of orbital magnetization'', emerged nearly one decade ago and is a logical continuation of the similar theory
of electric polarization:\cite{KSV,Resta} as the position operator $\textbf{r}$ is not well defined in the Bloch
representation, similar problem is anticipated for the orbital magnetization operator $(-e/2c)\textbf{r} \times \textbf{v}$,
which is also expressed through $\textbf{r}$.
Then,
the correct consideration of thermodynamic limit yielded a new and rather nontrivial expression for the orbital
magnetization, being another interesting manifestation of the
Berry-phase physics.\cite{Resta,XiaoPRL,ThonhauserPRL,CeresoliPRB,ShiPRL,ThonhauserIJMPB}

  The modern theory of the orbital magnetization is basically an one-electron theory. It does not say anything about
the form of exchange-correlation interactions.
Therefore, it would not be right to think that applications of the modern theory will automatically
resolve all previous problems, related to the form of the exchange-correlation functional and limitations of LSDA.

  Practical implementations of the modern theory of orbital magnetization are still rather limited.
Moreover, many of them are devoted to rather exotic
Haldane model Hamiltonian,\cite{Haldane} which is typically used in order to illustrate the
basic ideas (Refs.~\onlinecite{ThonhauserPRL,CeresoliPRB}) and to test computational schemes (Ref.~\onlinecite{CeresoliResta}).
The first-principles calculations were performed only for ferromagnetic metals Fe, Co, and Ni,
where the modern theory slightly improves the values of orbital magnetization
in comparison with the experimental data,\cite{ThonhauserIJMPB,fp3d}
and the orbital magnetoelectric coupling in insulators.\cite{mecoupling}

  At the same time, several important aspects of the modern theory remain obscure. To begin with, even if the previous
treatment of the orbital magnetization was incomplete,
it is not immediately clear what was missing in the ``standard'' expression (\ref{eqn:canonical})
and whether it can still be used in practical calculations for real materials. Then, what is the meaning of the
new corrections to Eq.~(\ref{eqn:canonical}), suggested by the modern theory?

  In this work, we apply the modern theory of the orbital magnetization to the series of representative distorted perovskite
transition-metal oxides
with the net ferromagnetic (FM)
moment in the ground state. Particularly, we consider
orthorhombic canted spin ferromagnet YTiO$_3$,
and three weak ferromagnets: orthorhombic LaMnO$_3$ and YVO$_3$, as well as monoclinic YVO$_3$.
These compounds differ by the type of the magnetic ground state
as well as the microscopic origin of the weak ferromagnetism:
regular spin canting caused by Dzyloshinskii-Moriya interactions in
orthorhombic systems (Refs.~\onlinecite{DM,Treves}) versus incomplete compensation of magnetic moments between two crystallographic
sublattices in monoclinic YVO$_3$.\cite{review2008} The magnetic structure of these materials depends on a subtle
interplay of the crystal distortion, relativistic SO coupling, and electron correlations in
the magnetically active bands. Therefore, from the computational point of view, it is more convenient to work with an
effective Hubbard-type model, derived from the first-principles electronic structure calculations,
and focusing on the behavior of these magnetically active bands.\cite{review2008}
The previous applications showed that such a strategy is very promising and the effective
model provides a reliable description for magnetic ground-state properties of
YTiO$_3$, YVO$_3$, and LaMnO$_3$.\cite{review2008,JPSJ,t2g}

  The rest of the paper is organized as follows. In Sec.~\ref{Sec:Th} we briefly remind to the reader the main
aspects of the modern theory of the orbital magnetization in solids.
In Sec.~\ref{sec:basis}, we identify the main contributions to the net orbital magnetic moment in the case
of basis -- when the Bloch wavefunction is expanded over localized Wannier-type orbitals, centered at magnetic sites.
Then, if the magnetic sites are located in the centers of inversion (the case that we consider), the net orbital
magnetic moment will have two contributions: the local one, which is given by the ``standard'' expression (\ref{eqn:canonical}),
and an ``itinerant'' correction to it, suggested by the modern theory. The behavior of the second part is closely related to that
of Chern invariant, which for the normal insulators with the canted FM structure can be viewed as a
``totally itinerant quantity'': the Chern invariant is given by certain Brilloin zone (BZ) integral. The individual
contributions to this integral in each $\textbf{k}$ can be finite.
However, the total integral, which can be regarded as a local (or site-diagonal) component
of some $\textbf{k}$-dependent property, is identically equal to zero. Then, in Sec.~\ref{sec:details}
we will briefly explain details of our calculations and in Sec.~\ref{Sec:Res} we will present numerical results for
YTiO$_3$, YVO$_3$, and LaMnO$_3$. We will show that the ``itinerant'' correction to the net orbital
magnetic moment is small. However, this small value is a result of cancelation of relatively large
contributions, coming from different parts of the BZ. Finally, in Sec. \ref{Sec:Con} we will summarize our work.

\section{\label{Sec:Th}General Theory}

  According to the modern theory of the orbital magnetization,\cite{ThonhauserPRL,CeresoliPRB,ShiPRL}
the net orbital magnetic moment of a normal periodic insulator
satisfies the following expression:
\begin{equation}
\boldsymbol{\cal M} = \frac{e}{2 \hbar c}
\mathrm{Im} \sum_{n}
\int_{BZ} \frac{d\textbf{k}}{\Omega}
\langle\partial_{\textbf{k}}u_{n\textbf{k}}|\times(H_{\textbf{k}}+E_{n\textbf{k}})|\partial_{\textbf{k}}u_{n\textbf{k}}\rangle,
\label{eqn:tot}
\end{equation}
where
$u_{n\textbf{k}}(\textbf{r})=e^{-i\textbf{k}\textbf{r}}\psi_{n\textbf{k}}(\textbf{r})$ is the cell-periodic eigenstate
of the Hamiltonian $H_{\textbf{k}}=e^{-i\textbf{k}\textbf{r}}He^{i\textbf{k}\textbf{r}}$,
corresponding to the eigenvalue $E_{n\textbf{k}}$, the
summation runs over occupied states, and the integration goes over the first BZ with the volume $\Omega$.
Eq.~(\ref{eqn:tot}) was
derived using different theoretical frameworks, including
semiclassical dynamics of Bloch electrons,\cite{XiaoPRL} the Wannier functions technique,\cite{ThonhauserPRL,CeresoliPRB}
and the perturbation theory in an external magnetic field.\cite{ShiPRL} It is important that all these methods
yield the same expression for $\boldsymbol{\cal M}$.

  In the modern theory of the orbital magnetization, the behavior of $\boldsymbol{\cal M}$ is closely related to
that of Chern invariants
\begin{equation}
\boldsymbol{\cal C} = - \frac{1}{2 \pi} \mathrm{Im} \sum_{n}
\int_{BZ}d\textbf{k} \, \langle \partial_{\textbf{k}}u_{n\textbf{k}} | \times | \partial_{\textbf{k}}u_{n\textbf{k}} \rangle,
\label{eqn:Chern_tot}
\end{equation}
which was originally introduced to characterize the Hall conductance.\cite{Thouless}
For the normal insulators, $\boldsymbol{\cal C}$ itself vanishes. Nevertheless, the integrand of Eq.~(\ref{eqn:Chern_tot})
(which is also related to the Berry curvature in the multi-band case) can be finite, depending on the symmetry of the crystal and the type
of the magnetic ground state. Thus, the finite value of $\boldsymbol{\cal M}$ in normal insulators can be viewed
as a result of additional modulation of the Berry curvature by the $\textbf{k}$-dependent quantities
$H_{\textbf{k}}$ and $E_{n\textbf{k}}$.

  Furthermore, it is understood that all electron-electron interactions are treated
in the spirit of Kohn-Sham DFT, that results in the self-consistent determination of the
single-particle Hamiltonian $H$ with the SO interaction.
It is important that the orbital magnetization (or related to it orbital currents) should
participate as an independent variable of the energy functional,
so that $\boldsymbol{\cal M}$ can be found through the expectation value of the
angular momentum operator in the basis of occupied Kohn-Sham orbitals $\psi_{n\textbf{k}}(\textbf{r})$ of the Hamiltonian $H$.\cite{Jansen}
Nevertheless, as was explained in the Introduction, the form of this functional is not known.
Therefore, in practical calculations, we have to rely on additional approximations. In the present work,
we use $H$ obtained in the mean-field Hartree-Fock (HF) approximation for the effective Hubbard-type model, which is derived
from the first-principles electronic structure calculations and is aimed to capture the behavior
of the magnetically active states near the Fermi level.\cite{review2008}
This model HF approach can be viewed as a functional of the
site-diagonal density matrix in the basis of localized Wannier orbitals,
which serve as the basis of the effective low-energy model.
Thus, the basic strategy of the present work is the following:
(i) The HF method is expected to reproduce the local part of the orbital moment,
which is
related to the site-diagonal density matrix by Eq.~(\ref{eqn:canonical});\cite{LDAU} and
(ii) We hope that it can also serve as a good starting point
for the analysis of other contributions to $\boldsymbol{\cal M}$.
Another possibility is to use current DFT, supplemented with some additional
approximations for the exchange-correlation energy.\cite{Ebert}

  The first term in Eq.~(\ref{eqn:tot}), which is called the ``local circulation'' $\boldsymbol{\cal M}^{LC}$,
is the lattice periodic contribution from the bulk Wannier orbitals, while the second terms
(the ``itinerant circulation'', $\boldsymbol{\cal M}^{IC}$) arises from the surface of the sample
and remains finite in the thermodynamic limit.\cite{ThonhauserPRL,CeresoliPRB} In the multi-orbital case,
each contribution become gauge invariant
(and, therefore, can be treated separately) if one uses the covariant derivatives:\cite{CeresoliPRB}
\begin{equation}
| \partial_{\textbf{k}}u_{n\textbf{k}} \rangle \rightarrow | \tilde{\partial}_{\textbf{k}}u_{n\textbf{k}} \rangle =
\left( 1-P_{\textbf{k}} \right) | \partial_{\textbf{k}}u_{n\textbf{k}} \rangle,
\label{eqn:CovariantD}
\end{equation}
where $P_{\textbf{k}} = \sum_n | u_{n\textbf{k}} \rangle \langle u_{n\textbf{k}} |$ is the ground-state projector.
The
total moment $\boldsymbol{\cal M} = \boldsymbol{\cal M}^{LC} + \boldsymbol{\cal M}^{IC}$ is not affected by the transformation (\ref{eqn:CovariantD}).
Moreover, in this covariant form, the formulation becomes gauge invariant not only for the BZ integrals, but also for their
integrants in each $\textbf{k}$-point of the reciprocal space.\cite{CeresoliPRB}
The same holds for the Chern invariants (\ref{eqn:Chern_tot}).
This allows us to discuss the $\textbf{k}$-dependence of the net orbital magnetic moments.

\section{\label{sec:basis}Orbital magnetization and basis}

  In this section we will consider how the main expression for $\boldsymbol{\cal M}$ [Eq.~(\ref{eqn:tot})]
can be reformulated
in the presence of basis. For these purposes, let us expand $|\psi_{n\textbf{k}}\rangle$ over some
basis of localized orbitals $|\phi_{\alpha}(\textbf{r}-\textbf{R})\rangle$,
centered at atomic sites $\textbf{R}$:
\begin{equation}
|\psi_{n\textbf{k}}\rangle = \frac{1}{\sqrt{N}}
\sum_{\alpha\textbf{R}}c^{\alpha}_{n\textbf{k}} e^{i\textbf{kR}}|\phi_{\alpha}(\textbf{r}-\textbf{R})\rangle,\\
\label{eqn:lmtofunc}
\end{equation}
where $N$ is the number of primitive cells, $\alpha$ is a combination of spin and orbital indices
(and, if necessary, the site indices in the primitive cell).
The basis itself satisfies the orthonormality condition:
\begin{equation}
\langle\phi_{\alpha'}(\textbf{r}-\textbf{R}')|\phi_{\alpha}(\textbf{r}-\textbf{R})\rangle=\delta_{\alpha'\alpha}\delta_{\textbf{R}'\textbf{R}}.
\label{eqn:orthogon}
\end{equation}
In our case, $\{ \phi_{\alpha}(\textbf{r}-\textbf{R}) \}$ is the basis of the Wannier functions, used for the
construction of the effective low-energy model.\cite{review2008}
However, it can be viewed in a more general sense: for example,
as the basis of nearly orthogonal linear muffin-tin orbitals of the LMTO method,\cite{LMTO}
or any orthonormal atomic-like basis.

  The use of the basis set is the general practice in numerical calculations.
However, apart from computational issues, the goal of this section is to understand what kind of
new contributions is provided by the modern theory of the orbital magnetization [Eq.~(\ref{eqn:tot})]
in comparison with the standard calculations, which are frequently formulated in the atomic-like
basis and based on the simplified expression (\ref{eqn:canonical}).\cite{OPB,Norman,LDAU} For these purposes,
we take the wavefunctions in the form (\ref{eqn:lmtofunc}) and substitute them in Eq.~(\ref{eqn:tot}). Then, the
\textbf{k}-space gradient of $| u_{n\textbf{k}} \rangle$ will have two contributions:
\begin{equation}
\begin{aligned}
|\partial_{\textbf{k}}u_{n\textbf{k}}\rangle&=-\frac{i}{\sqrt{N}}\sum_{\alpha\textbf{R}}(\textbf{r}-\textbf{R})e^{-i\textbf{k}(\textbf{r}-\textbf{R})}c^{\alpha}_{n\textbf{k}}|\phi_{\alpha}(\textbf{r}-\textbf{R})\rangle+ \\
&+\frac{1}{\sqrt{N}}\sum_{\alpha\textbf{R}}e^{-i\textbf{k}(\textbf{r}-\textbf{R})}\partial_{\textbf{k}}c^{\alpha}_{n\textbf{k}}|\phi_{\alpha}(\textbf{r}-\textbf{R})\rangle=|\partial_{\textbf{k}}u_{n\textbf{k}}\rangle^{\mathrm{I}}+|\partial_{\textbf{k}}u_{n\textbf{k}}\rangle^{\mathrm{II}},
\label{eqn:der}
\end{aligned}
\end{equation}
and we have to consider four possible contributions to Eq.~(\ref{eqn:tot}):
$\langle\partial_{\textbf{k}}u_{n\textbf{k}}|^{\mathrm{I}}\dots|\partial_{\textbf{k}}u_{n\textbf{k}}\rangle^{\mathrm{I}}$, $\langle\partial_{\textbf{k}}u_{n\textbf{k}}|^{\mathrm{I}}\dots|\partial_{\textbf{k}}u_{n\textbf{k}}\rangle^{\mathrm{II}}$, $\langle\partial_{\textbf{k}}u_{n\textbf{k}}|^{\mathrm{II}}\dots|\partial_{\textbf{k}}u_{n\textbf{k}}\rangle^{\mathrm{I}}$, and $\langle\partial_{\textbf{k}}u_{n\textbf{k}}|^{\mathrm{II}}\dots|\partial_{\textbf{k}}u_{n\textbf{k}}\rangle^{\mathrm{II}}$.
Moreover, we assume that all transition-metal sites are located in the inversion centers -- the situation, which is
indeed realized in perovskites with the $Pbnm$ and $P2_1/a$ structure.
Then, the Wannier functions
$\{ \phi_{\alpha}(\textbf{r} - \textbf{R}) \}$ will be either even or odd with respect to the inversion centers,
and we will have the following property:
\begin{equation}
\begin{aligned}
\langle\phi_{\alpha'}(\textbf{r}-\textbf{R}')|
\textbf{r}
|\phi_{\alpha}(\textbf{r}-\textbf{R})\rangle=\textbf{R}\delta_{\alpha'\alpha}\delta_{\textbf{R}'\textbf{R}}.
\label{eqn:Iproperty}
\end{aligned}
\end{equation}
In this case, after some tedious but rather straightforward algebra, which is explained in Supplemental Materials,\cite{SupM}
one can obtain the following expressions for the local circulation:
\begin{equation}
\begin{aligned}
\boldsymbol{\cal M}^{LC}&=\boldsymbol{\cal M}^0 + \Delta \boldsymbol{\cal M}^{LC}\\
&\equiv - \mu_{\rm B} \sum_{n} \sum_{\alpha \alpha'} \int_{BZ}
\,\frac{d\textbf{k}}{\Omega}
\,c^{\alpha'*}_{n\textbf{k}}\textbf{L}_{\textbf{k}}^{\alpha'\alpha}c^{\alpha}_{n\textbf{k}}\\
&+\frac{e}{2 \hbar c}\,\mathrm{Im}\sum_{n} \sum_{\alpha \alpha'}
\int_{BZ}\frac{d\textbf{k}}{\Omega}
\,\partial_{\textbf{k}}c^{\alpha'*}_{n\textbf{k}}\times H_{\textbf{k}}^{\alpha'\alpha}\partial_{\textbf{k}}c^{\alpha}_{n\textbf{k}},
\label{eqn:lcfin}
\end{aligned}
\end{equation}
and the itinerant circulation:
\begin{equation}
\begin{aligned}
\boldsymbol{\cal M}^{IC}&=\frac{e}{2 \hbar c}\,\mathrm{Im}\sum_{n} \sum_{\alpha}
\int_{BZ} \frac{d\textbf{k}}{\Omega} E_{n\textbf{k}}\,\partial_{\textbf{k}}c^{\alpha*}_{n\textbf{k}}\times\partial_{\textbf{k}}c^{\alpha}_{n\textbf{k}},
\label{eqn:icfin}
\end{aligned}
\end{equation}
where
\begin{equation}
H^{\alpha'\alpha}_{\textbf{k}}=\frac{1}{N}\sum\limits_{\textbf{R}\textbf{R}'}\,\langle\phi_{\alpha'}(\textbf{r}-\textbf{R}')|H|\phi_{\alpha}(\textbf{r}-\textbf{R})\rangle\,e^{i\textbf{k}(\textbf{R}-\textbf{R}')}
\label{eqn:lmtoham2}
\end{equation}
and
\begin{equation}
\textbf{L}_{\textbf{k}}^{\alpha'\alpha}=\frac{1}{N}\sum_{\textbf{R}\textbf{R}'}
\langle\phi_{\alpha'}(\textbf{r}-\textbf{R}') | (\textbf{r} - \textbf{R}')\times \textbf{p} |\phi_{\alpha}(\textbf{r}-\textbf{R})\rangle
e^{i\textbf{k}(\textbf{R}-\textbf{R}')}
\label{eqn:lbasis}
\end{equation}
are the Wannier matrix elements of Hamiltonian and periodic part of the angular momentum operator (divided by $\hbar$), respectively.
Moreover, Eq.~(\ref{eqn:lbasis}) implies that the momentum operator $\textbf{p}$ is related to the velocity $\textbf{v} = (i/\hbar) [H,\textbf{r}]$
in a ``nonrelativistic fashion'': $\textbf{p} = m\textbf{v}$.

  Thus, the local circulation has two terms. The first one ($\boldsymbol{\cal M}^0$) is the standard contribution,
that is given by periodic part of the angular momentum operator in the Wannier basis.
Due to orthonormality condition (\ref{eqn:orthogon}),
the main contributions in Eq.~(\ref{eqn:lbasis}) arise from the site-diagonal elements with
$\textbf{R} = \textbf{R}'$.
It can be best seen in the LMTO formulation,\cite{LMTO} where the tail of the basis function
$\phi_{\alpha}(\textbf{r}-\textbf{R})$ near the atomic site $\textbf{R}'$ is
expanded over
energy derivatives of $\{ \phi_{\alpha}(\textbf{r} - \textbf{R}') \}$. Then, since the function
is orthogonal to its energy derivative, all contributions with $\textbf{R} \neq \textbf{R}'$ in Eq.~(\ref{eqn:lbasis})
will vanish after the radial integration. Therefore, $\textbf{L}_{\textbf{k}}^{\alpha'\alpha}$ does not depend
on $\textbf{k}$ ($\textbf{L}_{\textbf{k}}^{\alpha'\alpha} \equiv \textbf{L}^{\alpha'\alpha}$), and $\boldsymbol{\cal M}^0$
is given by the standard expression,
$\boldsymbol{\cal M}^0 = - \mu_{\rm B} \mathrm{Tr}_\alpha  \{ \hat{\textbf{L}} \hat{\cal D}  \}$
in terms of the density matrix $\hat{\cal D} = [ {\cal D}^{\alpha \alpha'} ]$,
$$
{\cal D}^{\alpha \alpha'} = \sum_{n} \int_{BZ}
\, \frac{d\textbf{k}}{\Omega} \,c^{\alpha}_{n\textbf{k}} c^{\alpha'*}_{n\textbf{k}},
$$
where $\hat{\textbf{L}} \equiv [ \textbf{L}^{\alpha'\alpha} ]$ is the site-diagonal matrix and
$\mathrm{Tr}_\alpha$ is the trace over $\alpha$. Thus, the remaining term
$\Delta \boldsymbol{\cal M} = \Delta \boldsymbol{\cal M}^{LC} + \boldsymbol{\cal M}^{IC}$
can be viewed as a correction to $\boldsymbol{\cal M}^0$, suggested by the modern theory of the orbital magnetization.
$\Delta \boldsymbol{\cal M}$ has the same structure as Eq.~(\ref{eqn:tot}), and can be obtained
after replacing $| \partial_{\textbf{k}}u_{n\textbf{k}} \rangle$ by the column vector
$| \partial_{\textbf{k}}c_{n\textbf{k}} \rangle \equiv [ \partial_{\textbf{k}}c^{\alpha}_{n\textbf{k}} ]$
and $H_{\textbf{k}}$ by the matrix $\hat{H}_{\textbf{k}} \equiv [ H^{\alpha'\alpha}_{\textbf{k}} ]$ in the Wannier basis.
The same holds for the Chern invariants (\ref{eqn:Chern_tot}), where
$| \partial_{\textbf{k}}u_{n\textbf{k}} \rangle$ should be also replaced by
$| \partial_{\textbf{k}}c_{n\textbf{k}} \rangle$.

  In the following, we will also call $\boldsymbol{\cal M}^0$ the net \textit{local} magnetic moment and
$\Delta \boldsymbol{\cal M}$ the \textit{itinerant correction} to $\boldsymbol{\cal M}^0$.
This is because, for normal insulators, the Chern invariant itself can be regarded as a
totally itinerant quantity: It is given by the BZ integral of the Berry curvature.
The Berry curvature itself is $\textbf{k}$-dependent.
However, the local component of it, that is given by the BZ integration, is identically equal to zero.
Therefore, it is logical to view $\Delta \boldsymbol{\cal M}$, whose from is similar to $\boldsymbol{\cal C}$,
also as an itinerant contribution to the net orbital magnetic moment.
Moreover, for the totally localized states, $\hat{H}_{\textbf{k}}$ and $E_{n\textbf{k}}$ will not depend on $\textbf{k}$.
Therefore, in this case $\Delta \boldsymbol{\cal M}$ will vanish, similar to $\boldsymbol{\cal C}$.
This is another reason why $\Delta \boldsymbol{\cal M}$ can be associated with the itinerant contribution
to the orbital magnetic moment. One can also paraphrase this discussion in the following way:
the Berry curvature in the BZ integrals (\ref{eqn:lcfin})-(\ref{eqn:icfin}) acts as a ``filter'', which separates the local part of the
orbital magnetization from the itinerant one.

\section{\label{sec:details}Technical Details}

  All numerical calculations, reported in this work, have been performed for the effective low-energy model,
derived from the first-principles electronic structure calculations. First, we construct the effective Hubbard-type
model, describing the behavior of magnetically active $t_{2g}$ bands in the case of YTiO$_3$ and YVO$_3$,
and all $3d$ bands in the case of LaMnO$_3$.
For these purposes, we specify the basis of Wannier orbitals,
spanning the subspace of
these bands in the local-density approximation (LDA). Then, the parameters of crystal-field splitting,
SO interaction,
and transfer integrals of the effective model are given by the matrix elements of the LDA Hamiltonian
in the Wannier basis. The parameters of screened Coulomb and exchange interactions are obtained by
combining constrained LDA and
random-phase approximation (RPA) for the screening.\cite{review2008}
After the construction, the model was solved in the HF approximation. All details, including
the behavior of model parameters and results of HF calculations, can be found
in Refs.~\onlinecite{review2008,t2g,JPSJ}.

  Strictly speaking, if the model Hamiltonian $H$ includes the SO interaction term,
\begin{equation}
H_{\mathrm{SO}} = \frac{\hbar}{4m^2c^2}\boldsymbol{\sigma} \times \boldsymbol{\nabla}V \cdot \textbf{p},
\label{eqn:HSO}
\end{equation}
which originates from Pauli equations and is valid in the second order of $1/c$,
the velocity operator $\textbf{v}=(i/\hbar)[H,\boldsymbol{r}]$ will consists of two contributions:
\begin{equation}
\textbf{v}=\frac{\textbf{p}}{m}+\frac{\hbar}{4m^2c^2}\boldsymbol{\sigma}\times\boldsymbol{\nabla}V.
\label{eqn:velosity}
\end{equation}
The theory of orbital magnetization implies that
the second term in Eq.~(\ref{eqn:velosity}) can be neglected, that results in the
nonrelativistic expression $\textbf{p} = m \textbf{v}$. This can be done
because the contribution of the second term of Eq.~(\ref{eqn:velosity}) to the orbital magnetic moment
(\ref{eqn:tot}) is of the order of $1/c^3$, which is formally beyond the accuracy of Pauli equations.

  In order to calculate $\tilde{\partial}_{\textbf{k}}c_{n\textbf{k}}$ along the direction $i$ of the BZ,
we have used the discretized covariant derivative technique, which is well suited for insulators:\cite{CeresoliPRB,ThonhauserIJMPB}
\begin{equation}
\tilde{\partial}_ic_{n\textbf{k}}=\frac{1}{2|\textbf{q}|}\left(\tilde{c}_{n\textbf{k}+\textbf{q}}-\tilde{c}_{n\textbf{k}-\textbf{q}}\right),
\end{equation}
where $\textbf{q}$ is the vector that connects $\textbf{k}$ with the nearby point in the direction $i$ and
$\tilde{c}_{n\textbf{k}+\textbf{q}}$ is
the ``dual'' state, defined in terms of the overlap matrix
$(S_{\textbf{k},\textbf{k}+\textbf{q}})_{nn'}=\langle c_{n\textbf{k}} | c_{n'\textbf{k}+\textbf{q}} \rangle$ as
\begin{equation}
\tilde{c}_{n\textbf{k}+\textbf{q}}=\sum_{n'}(S^{-1}_{\textbf{k},\textbf{k}+\textbf{q}})_{n'n}c_{n'\textbf{k}}.
\end{equation}

  As for the $\textbf{k}$-space integration, we have used the grid of about
$70$$\times$$70$$\times$$50$ points in the first BZ, which guarantees
an excellent convergence for $\Delta \boldsymbol{\cal M}$ depending on the number of $\textbf{k}$-points.\cite{SupM}

\section{\label{Sec:Res}Results}

\subsection{\label{YTiO3}YTiO$_3$}

\begin{figure}
\begin{center}
\includegraphics[width=8cm]{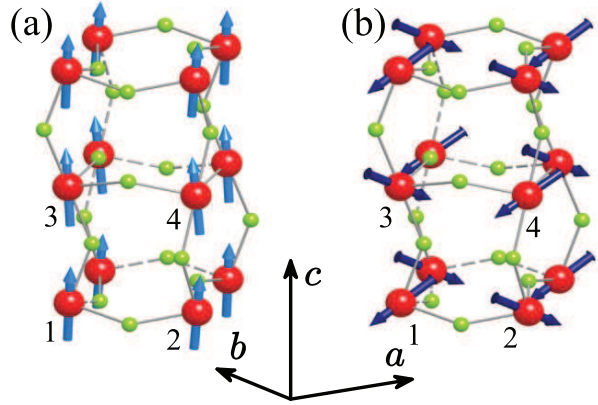}
\end{center}
\caption{\label{fig.YTiO3HF}(Color online)
Distribution of spin (a) and orbital (b) magnetic moments as obtained in the mean-field
Hartree-Fock calculations for the low-energy model of YTiO$_3$.\cite{t2g}
The titanium atoms are indicated by the big red (dark) spheres and the oxygen atoms
are indicated by the small green (grey) spheres. For the sake of clarity,
the arrows for the orbital magnetic moments were scaled in order to have
the same length as for the spin magnetic moments.
}
\end{figure}

  YTiO$_3$ crystallizes in the orthorhombic $Pbnm$ structure (in our calculations, we used the
experimental structure parameters, measured at 2~K).\cite{YTiO3exp} Below $T_{\rm C} \approx$ 29~K, it forms the canted
FM structure, where the net FM
moment is parallel to the orthorhombic $\boldsymbol{c}$ axis. Two other components of the magnetic moments,
parallel to the orthorhombic $\boldsymbol{a}$ and $\boldsymbol{b}$ axes, are ordered antiferromagnetically.
The type of this ordering is G and A, respectively.
Such magnetic structure can be abbreviated as
G$_{\boldsymbol{a}}$-A$_{\boldsymbol{b}}$-F$_{\boldsymbol{c}}$.
It was successfully reproduced by our mean-field HF approximation for the low-energy model.
The details of these calculations can be found in Ref.~\onlinecite{t2g}
and the
obtained magnetic structure is summarized in Fig.~\ref{fig.YTiO3HF}.
In this case, the vector of the spin magnetic moment at the site $1$ is
$(-$$0.021, -$$0.127, \, 0.986)$ $\mu_{\rm B}$ and the vector $\boldsymbol{\mu}^0$ of orbital magnetic moment is
$(-$$0.033, -$$0.001, -$$0.018)$ $\mu_{\rm B}$. Therefore, the net local orbital magnetic moment ${\cal M}^0_c$
(per one primitive cell of YTiO$_3$, containing four Ti atoms) is $-$$0.072$ $\mu_{\rm B}$ (Table~\ref{tab.L}).
As was explained above, it is parallel to the $\boldsymbol{c}$ axis.
\begin{table}[h!]
\caption{\label{tab.L}Different contributions to the net orbital magnetic moment, as obtained in the mean-field Hartree-Fock calculations
for the low-energy model: the local moment $\boldsymbol{\cal M}^0$, given by periodic part of the orbital momentum operator
in the Wannier basis, and two itinerant contributions, due to the local and itinerant circulation
($\Delta \boldsymbol{\cal M}^{LC}$ and $\boldsymbol{\cal M}^{IC}$, respectively).
All values are in $\mu_\textrm{B}$ per one primitive cell, containing four transition-metal sites.}
\begin{ruledtabular}
\begin{tabular}{lcccccc}
Compound & Direction & $\boldsymbol{\cal M}^0$ & $\Delta \boldsymbol{\cal M}^{LC}$ & $\boldsymbol{\cal M}^{IC}$
& $\Delta \boldsymbol{\cal M}^{LC} + \boldsymbol{\cal M}^{IC}$  \\
\hline
YTiO$_{3}$ ($Pbnm$) & $||\boldsymbol{c}$ & $-0.072$ & $-1.22\cdot10^{-5}$ & $\phantom{-}2.63\cdot10^{-4}$ & $\phantom{-}2.50\cdot10^{-4}$ \\
LaMnO$_{3}$ ($Pbnm$) & $||\boldsymbol{c}$ & $-0.032$ & $\phantom{-}1.05\cdot10^{-4}$ & $-2.28\cdot10^{-4}$ & $-1.23\cdot10^{-4}$ \\
YVO$_{3}$ ($Pbnm$) & $||\boldsymbol{a}$ & $-0.004$ & $-6.75\cdot10^{-4}$ & $-1.30\cdot10^{-4}$ & $-8.05\cdot10^{-4}$ \\
YVO$_{3}$ ($P2_1/a$) & $||\boldsymbol{b}$ & $-0.020$  & $\phantom{-}3.30\cdot10^{-6}$ & $-2.29\cdot10^{-5}$ & $-1.96\cdot10^{-5}$ \\
\end{tabular}
\end{ruledtabular}
\end{table}

  Then, we evaluate the itinerant correction $\Delta \boldsymbol{\cal M}$, resulting from the local and itinerant circulation terms.
These results are summarized in Table~\ref{tab.L}.
As expected, the projections of $\Delta \boldsymbol{\cal M}$ onto the orthorhombic $\boldsymbol{a}$ and $\boldsymbol{b}$ axes
are identically equal to zero.
The $\boldsymbol{c}$ projection ($\Delta {\cal M}_c$) is finite. However, it
is more than two orders of magnitude smaller than ${\cal M}^0_c$ and, therefore, can be safely neglected.
In principle, this result is also anticipated for the considered transition-metal oxides, which are frequently regarded as
Mott insulators and in which the magnetically active $3d$ states are relatively well localized.

\begin{figure}
\begin{center}
\includegraphics[width=10cm]{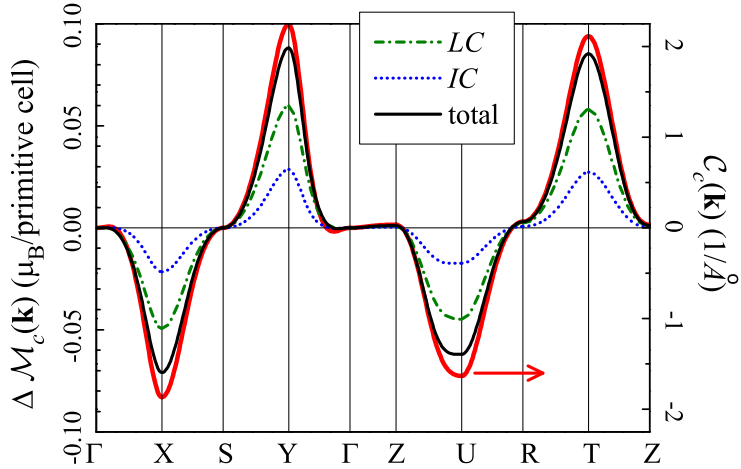}
\end{center}
\caption{\label{fig.YTiO3k}(Color online)
Behavior of
itinerant
contributions to the net orbital magnetic moment in YTiO$_3$ (left axis) and corresponding Chern invariant (right axis)
in the reciprocal space, along high-symmetry directions of the
Brillouin zone.
Two partial contributions to the net orbital moment,
associated with the
local ($\Delta \boldsymbol{\cal M}^{LC}$) and itinerant ($\boldsymbol{\cal M}^{IC}$) circulation are denoted as $LC$ and $IC$, respectively,
and the sum of these two contributions is denoted as `total'.
}
\end{figure}

\begin{figure}
\begin{center}
\includegraphics[height=10cm]{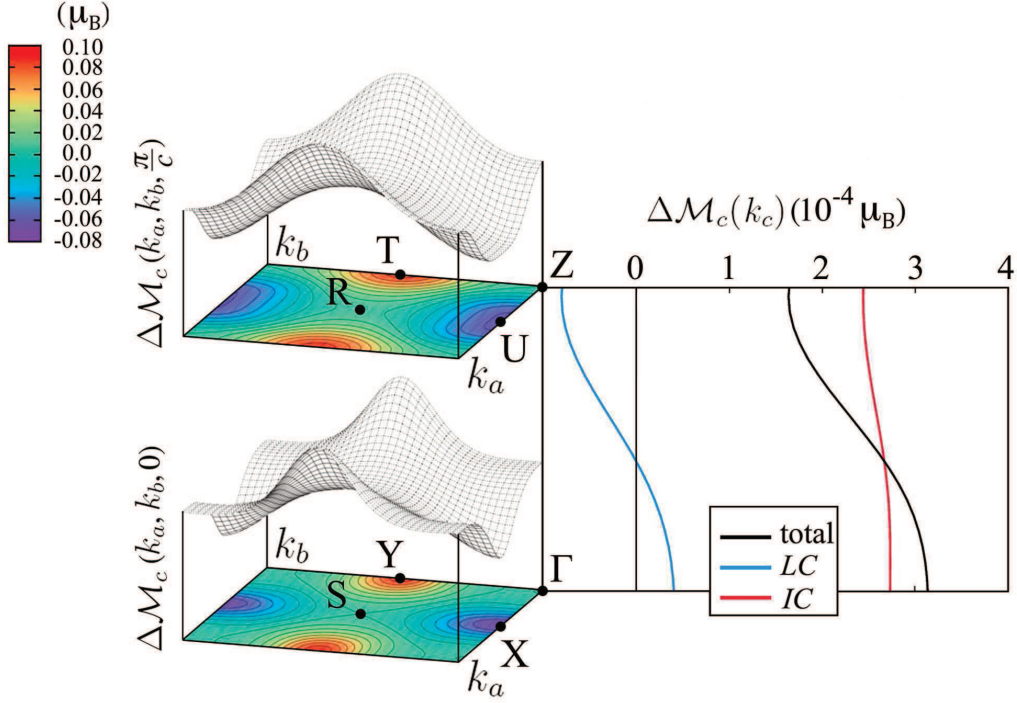}
\end{center}
\caption{\label{YTiO3_3D}
Left panel: Three-dimensional plot of $\Delta {\cal M}_c (\textbf{k}) \equiv \Delta {\cal M}_c (k_a,k_b,k_c)$ for $k_c = 0$ and $\frac{1}{2}$
in the case of YTiO$_3$.
Right panel: the $k_c$-dependence of $\Delta {\cal M}_c (k_c)$, obtained after the integration of $\Delta {\cal M}_c (\textbf{k})$
over $k_a$ and $k_b$, and its partial contributions associated with local and itinerant circulation
terms ($LC$ and $IC$, respectively).
}
\end{figure}

  Nevertheless, it is interesting to
gain some insight by investigating the origin of such a small value. For these purposes we analyze the integrand
$$
\Delta \boldsymbol{\cal M} (\textbf{k}) = \frac{e}{2 \hbar c} \, \mathrm{Im} \sum_{n}
\langle \partial_{\textbf{k}}c_{n\textbf{k}} |\times(\hat{H}_{\textbf{k}}+E_{n\textbf{k}})| \partial_{\textbf{k}}c_{n\textbf{k}} \rangle
$$
of
$$
\Delta \boldsymbol{\cal M} =  \int_{BZ} \frac{d\textbf{k}}{\Omega} \, \Delta \boldsymbol{\cal M} (\textbf{k})
$$
and plot it
along high-symmetry directions of the BZ (see Fig.~\ref{fig.YTiO3k}).
Notations of the high-symmetry points of the BZ were taken from the book of Bradley and Cracknell.\cite{BradlayCracknell}
We obtained that two components, $\Delta {\cal M}_a (\textbf{k})$ and $\Delta {\cal M}_b (\textbf{k})$,
are identically equal to zero in each $\textbf{k}$-point, while $\Delta {\cal M}_c (\textbf{k})$ can be finite and, moreover,
strongly depend on $\textbf{k}$. This behavior is consistent with the G$_{\boldsymbol{a}}$-A$_{\boldsymbol{b}}$-F$_{\boldsymbol{c}}$
symmetry of the magnetic ground state.\cite{remark1}
$\Delta {\cal M}_c (\textbf{k})$ reaches its maximal value of $0.088$~$\mu_\textrm{B}$ in the point
$\textrm{Y} = (0,\frac{1}{2},0)$ of the BZ (in units of reciprocal lattice translations),
which is comparable with ${\cal M}^0_c$. Thus, the individual contributions
$\Delta {\cal M}_c (\textbf{k})$ can be large. However, there is also a large
cancelation between
positive and negative contributions to $\Delta {\cal M}_c$ around the $\textrm{Y}$ and
$\textrm{X} = (\frac{1}{2},0,0)$ points, respectively. Similar situation occurs at the BZ boundary $k_c = \frac{1}{2}$, where again
the large positive contribution around $\textrm{T} = (0,\frac{1}{2},\frac{1}{2})$
is nearly canceled by the negative contribution
around $\textrm{U} = (\frac{1}{2},0,\frac{1}{2})$. This result is summarized in Fig.~\ref{YTiO3_3D},
where we plot
$\Delta {\cal M}_c (k_a,k_b,k_c)$ for $k_c = 0$ and $\frac{1}{2}$, as well as the integrated value
$$
\Delta {\cal M}_c (k_c) = \frac{|\boldsymbol{a}| |\boldsymbol{b}|}{4 \pi^2}
\int_{-\pi/a}^{\pi/a} d k_a \int_{-\pi/b}^{\pi/b} d k_b \, \Delta {\cal M}_c (k_a,k_b,k_c).
$$
One can clearly see that $\Delta {\cal M}_c (k_a,k_b,k_c)$ only weakly depends on $k_c$. For each $k_c$,
there is a strong cancelation of the positive and negative contributions to $\Delta {\cal M}_c (k_a,k_b,k_c)$,
arising from
$\textbf{k} = (0,\frac{1}{2},k_c)$ and $(\frac{1}{2},0,k_c)$, respectively. This cancelation readily explains
the small value of $\Delta {\cal M}_c (k_c)$. Finally, the integration of $\Delta {\cal M}_c (k_c)$ over $k_c$ yields
the total value of $\Delta {\cal M}_c$, reported in Table~\ref{tab.L}. Thus, the small value of $\Delta {\cal M}_c$
is the result of strong cancelation of relatively large contributions $\Delta {\cal M}_c (\textbf{k})$, coming from
different parts of the BZ. This is the reason why we consider $\Delta \boldsymbol{\cal M}$ as an itinerant quantity.
Moreover, the strong $\textbf{k}$-dependence of $\Delta \boldsymbol{\cal M}$ implies that
after the Fourier transformation to the real space,
in addition to the small site-diagonal component, this quantity will have a large nonlocal
(or off-diagonal with respect to the atomic sites) part.
Since the $\textbf{k}$-dependence is smooth, this Fourier series will converge and such a
real-space analysis can be justified.

  As was already pointed out in Sec.~\ref{Sec:Th},
this behavior is closely related to that of the Chern invariants.
For our purposes, it is convenient to rewrite $\boldsymbol{\cal C}$ in the following form:
$$
\boldsymbol{\cal C} = \frac{1}{\Omega} \int_{BZ}d\textbf{k} \, \boldsymbol{\cal C} (\textbf{k}),
$$
where
$$
\boldsymbol{\cal C} (\textbf{k}) = - \frac{\Omega}{2 \pi} \mathrm{Im} \sum_{n}
\langle \partial_{\textbf{k}}c_{n\textbf{k}} | \times | \partial_{\textbf{k}}c_{n\textbf{k}} \rangle.
$$
For the normal insulators, $\boldsymbol{\cal C}$ is zero, and this property is perfectly reproduced
by our calculations. However,
due to the specific symmetry of the G$_{\boldsymbol{a}}$-A$_{\boldsymbol{b}}$-F$_{\boldsymbol{c}}$
ground state of YTiO$_3$,\cite{remark1}
the integrand ${\cal C}_c (\textbf{k})$ can be
finite in the individual $\textbf{k}$-points,
while two other projections of $\boldsymbol{\cal C} (\textbf{k})$ onto the orthorhombic
$\boldsymbol{a}$ and $\boldsymbol{b}$ axes are identically equal to zero.
Furthermore, the $\textbf{k}$-dependence of ${\cal C}_c (\textbf{k})$ is very close to that of $\Delta {\cal M}_c (\textbf{k})$
(see Fig.~\ref{fig.YTiO3k}). Thus, in the case of Chern invariant ${\cal C}_c (\textbf{k})$,
the contributions from different parts of the BZ exactly cancel each other. However, in the expression for
$\Delta {\cal M}_c$, the $\textbf{k}$-dependence of ${\cal C}_c (\textbf{k})$ for each band is additionally modulated with
$\textbf{k}$-dependent quantities $\hat{H}_{\textbf{k}}$ and $E_{n\textbf{k}}$, that leads to a small but finite
value of $\Delta {\cal M}_c$ (see Table~\ref{tab.L}). It also explains why $\Delta {\cal M}^{LC}_c (\textbf{k})$
and ${\cal M}^{IC}_c (\textbf{k})$ reveal very similar $\textbf{k}$-dependence: in both cases, it is
dictated by the $\textbf{k}$-dependence of ${\cal C}_c (\textbf{k})$, which appears to be more fundamental quantity.

\subsection{\label{LaMnO3}LaMnO$_3$}

\begin{figure}
\begin{center}
\includegraphics[width=8cm]{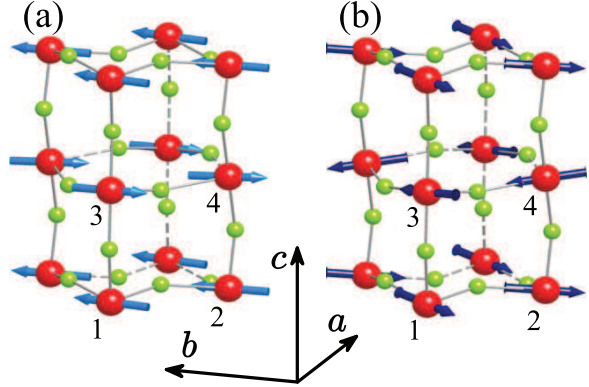}
\end{center}
\caption{\label{fig.LaMnO3HF}(Color online)
Distribution of spin (a) and orbital (b) magnetic moments as obtained in the mean-field
Hartree-Fock calculations for the low-energy model of LaMnO$_3$.
The manganese atoms are indicated by the big red (dark) spheres and the oxygen atoms
are indicated by the small green (grey) spheres. For the sake of clarity,
the arrows for the orbital magnetic moments were scaled in order to have
the same length as for the spin magnetic moments.
}
\end{figure}

  LaMnO$_3$ is another compound, crystallizing in the orthorhombic $Pbnm$ structure.\cite{LaMnO3exp}
It has the same G$_{\boldsymbol{a}}$-A$_{\boldsymbol{b}}$-F$_{\boldsymbol{c}}$
type of the magnetic ground state,
which is realized below $T_{\rm N} \approx$ 140~K.\cite{Matsumoto}
This magnetic ground state was successfully reproduced in our mean-field HF calculations
for the low-energy model.
The basic difference from YTiO$_3$
is that the spin magnetic structure is nearly A-type antiferromagnetic (AFM)
and the FM canting of spins in the $\boldsymbol{c}$ direction
is really small. In this sense, LaMnO$_3$ is the canonical weak ferromagnet. Nevertheless, the
orbital magnetic structure is strongly deformed:
in comparison with the spin one,
there is a large deviation from the collinear A-type AFM alignment
and an appreciable canting of the orbital magnetic moments
in the direction of $\boldsymbol{a}$ and $\boldsymbol{c}$,
which can be seen even visually in Fig.~\ref{fig.LaMnO3HF}.
The vector of spin magnetic moment at the site 1 is
$(\, 0.354, \, 3.952, \, 0.111)$ $\mu_{\rm B}$ and the one of orbital magnetic moment $\boldsymbol{\mu}^0$ is
$(-$$0.030, -$$0.057, -$$0.008)$ $\mu_{\rm B}$. Thus, the net orbital magnetic moment $\boldsymbol{\cal M}^0$
is $-$$0.032$ $\mu_{\rm B}$ (Table~\ref{tab.L}).

\begin{figure}
\begin{center}
\includegraphics[width=10cm]{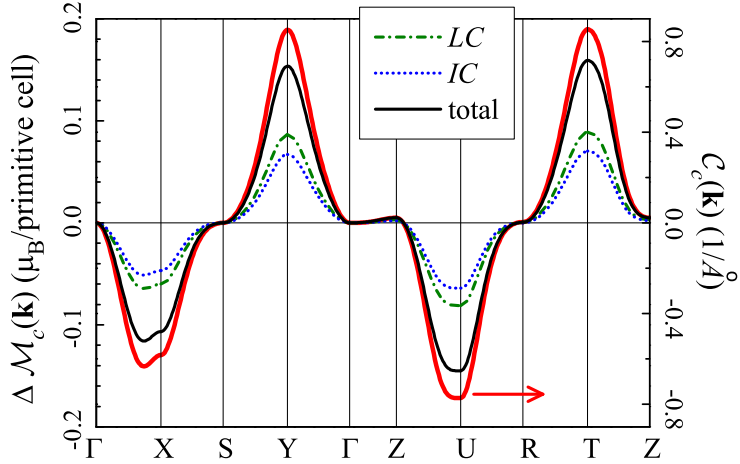}
\end{center}
\caption{\label{fig.LaMnO3k}(Color online)
Behavior of
itinerant
contributions to the net orbital magnetic moment in LaMnO$_3$ (left axis) and corresponding Chern invariant (right axis)
in the reciprocal space, along high-symmetry directions of the
Brillouin zone.
Two partial contributions to the net orbital moment,
associated with the
local ($\Delta \boldsymbol{\cal M}^{LC}$) and itinerant ($\boldsymbol{\cal M}^{IC}$) circulation are denoted as $LC$ and $IC$, respectively,
and the sum of these two contributions is denoted as `total'.
}
\end{figure}

  The behavior of $\Delta \boldsymbol{\cal M} (\textbf{k})$ is qualitatively the same as
in YTiO$_3$: it has similar structure and similar type of cancelation between different
parts of the BZ (Fig.~\ref{fig.LaMnO3k}).
Taking into account that YTiO$_3$ and LaMnO$_3$ have the same type of crystal structure and the magnetic ground state,
such similarity is not surprising.
The main difference is the magnitude of the effect, which is much more pronounced in LaMnO$_3$:
the values of $\Delta {\cal M}_c (\textbf{k})$ in the $\textrm{Y}$ and $\textrm{T}$ points
are $0.154$ $\mu_{\rm B}$ and $0.159$ $\mu_{\rm B}$, respectively,
which exceed ${\cal M}^0_c$ by factor five. However, there is again a strong cancelation with the negative
contributions around the $\textrm{X}$ and $\textrm{U}$ points of the BZ, which, after the integration, leads to the
small value of $\Delta {\cal M}_c$. Moreover, in LaMnO$_3$ there is a partial cancelation between
$LC$ and $IC$ contributions to $\Delta {\cal M}_c$ (see Table~\ref{tab.L}).

  Like in YTiO$_3$, the $\textbf{k}$-dependence of $\Delta {\cal M}_c (\textbf{k})$ in LaMnO$_3$
follows the form of ${\cal C}_c (\textbf{k})$ (Fig.~\ref{fig.LaMnO3k}).
Nevertheless, one interesting aspect is that
the amplitude of
${\cal C}_c (\textbf{k})$ in LaMnO$_3$ is smaller than in YTiO$_3$ (Fig.~\ref{fig.YTiO3k}), while for
$\Delta {\cal M}_c (\textbf{k})$ the situation is exactly the opposite.
This difference may be related to the number of occupied bands
(16 in the case of LaMnO$_3$ versus 4 in the case of YTiO$_3$). Thus, the amplitude of $\Delta {\cal M}_c (\textbf{k})$
may be larger in LaMnO$_3$ because the number of occupied bands is larger.
Another possibility is that
the contributions of different bands cancel each other and this cancelation occurs in a different way in the case of
${\cal C}_c (\textbf{k})$ and $\Delta {\cal M}_c (\textbf{k})$.

\subsection{\label{YVO3}YVO$_3$}

\begin{figure}
\begin{center}
\includegraphics[width=8cm]{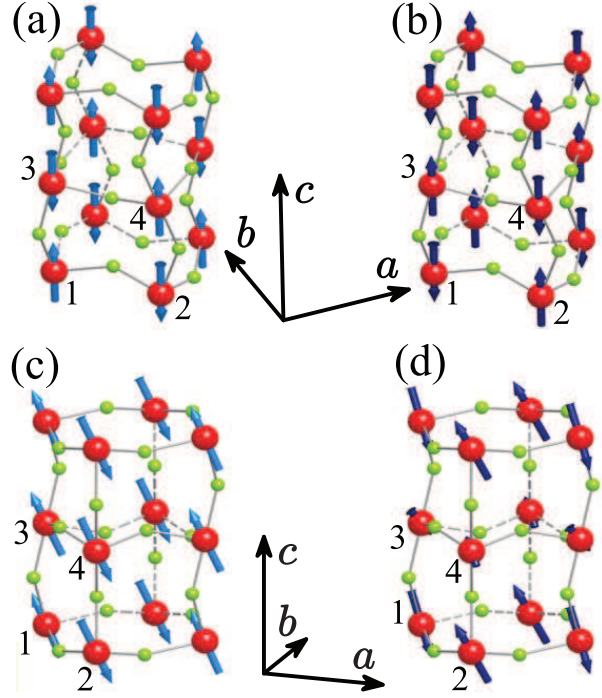}
\end{center}
\caption{\label{fig.YVO3HF}(Color online)
Distribution of spin (a and c) and orbital (b and d) magnetic moments as obtained in the mean-field
Hartree-Fock calculations for the low-energy model of YVO$_3$ in the
orthorhombic (a and b) and monoclinic (c and d) phases.
The vanadium atoms are indicated by the big red (dark) spheres and the oxygen atoms
are indicated by the small green (grey) spheres. For the sake of clarity,
the arrows for the orbital magnetic moments in the sublattice (1,2) were scaled in order to have
the same length as for the spin magnetic moments.
The orbital magnetic moments in the sublattice (3,4) are additionally quenched
by stronger crystal field.
}
\end{figure}

  YVO$_3$ has two crystallographic modifications: orthorhombic $Pbnm$, which is realized below 77~K,
and monoclinic $P2_1/a$ above 77~K (in our calculations, we use the experimental structure parameters
at 65~K and 100~K, respectively).\cite{YVO3exp}
The magnetic structure, realized in the orthorhombic $Pbnm$ phase is
F$_{\boldsymbol{a}}$-C$_{\boldsymbol{b}}$-G$_{\boldsymbol{c}}$ (Fig.~\ref{fig.YVO3HF}).
According to the mean-field HF calculations
for the low-energy model, the vector of spin magnetic moment at the site 1 is
$(-$$0.016, \, 0, \, 1.969)$ $\mu_{\rm B}$ and the vector of orbital magnetic moment is
$(-$$0.001, \, 0.001, -$$0.186)$  $\mu_{\rm B}$. Thus, the $\boldsymbol{c}$ projection
clearly dominates, while two other projections
are substantially smaller. The net orbital magnetic moment is only $-$$0.004$ $\mu_{\rm B}$,
which is parallel to the orthorhombic $\boldsymbol{a}$ axis.

  The monoclinic phase of YVO$_3$ has two inequivalent pairs of V sites,
which are denoted  in Fig.~\ref{fig.YVO3HF} as (1,2) and (3,4).
Within each pair, the $\boldsymbol{a}$ and $\boldsymbol{c}$ projections of the magnetic moments
are coupled antiferromagnetically, while the $\boldsymbol{b}$ projection is ferromagnetic.
According to the mean-field HF calculations for the low-energy model,
the vectors of spin magnetic moments at the sites 1 and 3
are $(-$$0.850, \, 0.077, \, 1.785)$ $\mu_{\rm B}$ and $(-$$0.875, -$$0.032, \, 1.764)$ $\mu_{\rm B}$, respectively,
and the vectors of orbital magnetic moments are
$(\, 0.074, -$$0.046, -$$0.173)$ $\mu_{\rm B}$ and
$(\, 0.043, \, 0.036, -$$0.073)$ $\mu_{\rm B}$, respectively.
The local orbital magnetic moments in the sublattice (3,4) are substantially smaller due to
additional quenching by stronger crystal field (see Ref.~\onlinecite{t2g} for details).
Thus, there is a partial cancelation
of the FM magnetization between two sublattices. However,
due to the additional quenching in the sublattice (3,4),
this cancelation is not complete and the system remains weakly ferromagnetic.
The net orbital magnetic moment $\boldsymbol{\cal M}^0$
is $-$$0.02$ $\mu_{\rm B}$, which is parallel to the monoclinic $\boldsymbol{b}$ axis.
The directions of the net magnetic moment and, therefore, the type of the magnetic ground state in the
orthorhombic and monoclinic phases are well consistent with the experimental data.\cite{Ren}

\begin{figure}
\begin{center}
\includegraphics[width=10cm]{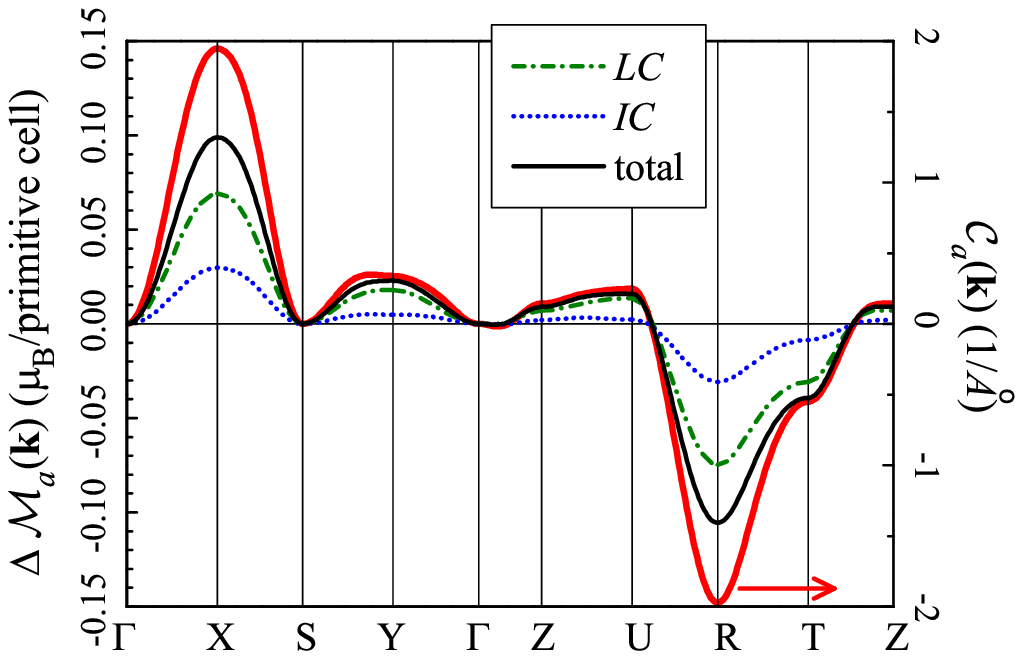}
\end{center}
\caption{\label{fig.YVO3ok}(Color online)
Behavior of
itinerant
contributions to the net orbital magnetic moment in orthorhombic YVO$_3$ (left axis) and corresponding Chern invariant (right axis)
in the reciprocal space, along high-symmetry directions of the
Brillouin zone.
Two partial contributions to the net orbital moment,
associated with the
local ($\Delta \boldsymbol{\cal M}^{LC}$) and itinerant ($\boldsymbol{\cal M}^{IC}$) circulation are denoted as $LC$ and $IC$, respectively,
and the sum of these two contributions is denoted as `total'.
}
\end{figure}

  The type of the magnetic ground state in orthorhombic YVO$_3$ is different from the one of YTiO$_3$ and LaMnO$_3$.
As a result, the $\textbf{k}$-dependence of $\boldsymbol{\cal C} (\textbf{k})$ and
$\Delta \boldsymbol{\cal M} (\textbf{k})$ is also different. Since the
net magnetic moment is parallel to the orthorhombic $\boldsymbol{a}$ axis, only $\boldsymbol{a}$ projection of
$\Delta \boldsymbol{\cal M}$ is finite, while two other projections are identically equal to zero.
Then, $\Delta {\cal M}_a(\textbf{k})$ reaches the maximal value of $0.099$ $\mu_{\rm B}$ in the $\textrm{X}$ point of the BZ (Fig.~\ref{fig.YVO3ok}),
which exceeds the net local magnetic moment ${\cal M}^0_a$ by more than one order of magnitude (Table~\ref{tab.L}).
There are other positive contributions, originating from the $\textrm{X}$, $\textrm{Z} = (0,0,\frac{1}{2})$, and
$\textrm{U}$ points of the BZ. Nevertheless, they are well compensated by the negative contributions, coming from the $\textrm{T}$
and $\textrm{R} = (\frac{1}{2},\frac{1}{2},\frac{1}{2})$ points of the BZ,
that again results in the small value of $\Delta {\cal M}_a$ (Table~\ref{tab.L}).
This behavior is totally consistent with the form of ${\cal C}_a (\textbf{k})$.

\begin{figure}
\begin{center}
\includegraphics[width=10cm]{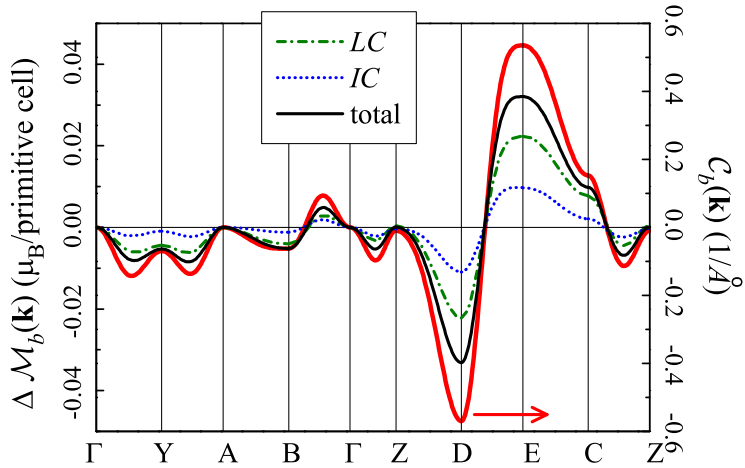}
\end{center}
\caption{\label{fig.YVO3mk}(Color online)
Behavior of
itinerant
contributions to the net orbital magnetic moment in monoclinic YVO$_3$ (left axis) and corresponding Chern invariant (right axis)
in the reciprocal space, along high-symmetry directions of the
Brillouin zone.
Two partial contributions to the net orbital moment,
associated with the
local ($\Delta \boldsymbol{\cal M}^{LC}$) and itinerant ($\boldsymbol{\cal M}^{IC}$) circulation are denoted as $LC$ and $IC$, respectively,
and the sum of these two contributions is denoted as `total'.
}
\end{figure}

  A completely different type of cancelation occurs in the monoclinic phase of YVO$_3$.
In this case, the net orbital moment
is parallel to the monoclinic $\boldsymbol{b}$ axis (Table~\ref{tab.L}), and  $\Delta {\cal M}_b (\textbf{k})$ has the
largest magnitude in the plane $k_c = \frac{1}{2}$, where the region of positive values
around the point $\textrm{E} = (\overline{\frac{1}{2}},\frac{1}{2},\frac{1}{2})$ is nearly canceled by the region of negative values around the point
$\textrm{D} = (\overline{\frac{1}{2}},0,\frac{1}{2})$ (Fig.~\ref{fig.YVO3mk}).
This behavior is again consistent with the form of ${\cal C}_b (\textbf{k})$ and explains the small value of
integrated $\Delta {\cal M}_b$ in Table~\ref{tab.L}.

\section{\label{Sec:Con}Conclusions}

  We have applied the modern theory of orbital magnetization to the series of characteristic distorted perovskite
transition-metal oxides
with a net FM moment in the ground state.
Our applications cover the examples of canted (but yet robust) ferromagnetism in orthorhombic YTiO$_3$ as well as
weak ferromagnetism caused by either
antisymmetric Dzyalishinskii-Moriya interactions in orthorhombic LaMnO$_3$ and YVO$_3$ or
imperfect cancelation of magnetic moments between two crystallographic sublattices in monoclinic YVO$_3$.
Our numerical calculations suggest that, for all these
compounds, the orbital magnetization can be well described by the ``standard'' expression (\ref{eqn:canonical}),
in terms of the angular momentum operator and the site-diagonal density matrix, while all the ``itinerant'' corrections,
originating from the modern theory, are negligibly small.
Nevertheless, the smallness of these corrections
is the result of
rather nontrivial cancelation of relatively large contributions coming from different parts of the BZ.

  There is a big difference
in the behavior of orbital magnetization and ferroelectric (FE) polarization
in improper multiferroics.
In the latter case, the inversion symmetry is broken by some complex magnetic order, while the crystal structure itself,
to a good approximation, can be regarded as centrosymmetric. Then, if the magnetic sites are located in
the centers of inversion, Eq.~(\ref{eqn:Iproperty}) yields
$\langle\phi_{\alpha'}(\textbf{r}-\textbf{R})|
\textbf{r}-\textbf{R}
|\phi_{\alpha}(\textbf{r}-\textbf{R})\rangle = 0$, which means that there is no ``local FE polarization'',
associated with the basis functions of the magnetic sites. Finite value of the FE polarization in this case
is related to the
$\textbf{k}$-dependence of the coefficients $\{ c_{n\textbf{k}} \}$ of expansion
of the Bloch eigenfunctions over the basis functions
and can be obtained by applying the Berry-phase theory only for $\{ c_{n\textbf{k}} \}$.\cite{FE_model,PRB13}
In this sense, and using an analogy with the modern theory for the orbital magnetization, one can say that
the FE polarization in improper multiferroics is entirely itinerant quantity and can even serve as the measure of itineracy
of magnetic system.\cite{PRB13}

  The behavior of orbital magnetization in the normal FM insulators is fundamentally different.
In this case, there are finite local magnetic moments, which are expressed in terms of matrix elements of the
angular momentum operator in the Wannier basis, and these local magnetic moments provide the main contribution to the
net orbital magnetic moment. The itinerant corrections $\Delta \boldsymbol{\cal M}$ to this net FM moment,
originating from the $\textbf{k}$-dependence of $\{ c_{n\textbf{k}} \}$, are considerably smaller. Thus, the
orbital magnetic moment is mainly a local quantity.

  The form of $\boldsymbol{\cal M}(\textbf{k})$ in the reciprocal space follows
the behavior of Chern invariants. Although the full integral over the BZ is small (or identically
equals to zero in the case
of Chern invariants), the integrand itself is finite and, moreover,
can be
strongly $\textbf{k}$-dependent. By tracing this discussion back to the real space by means of
the Fourier transform, this would
mean that the considered quantities will have nonlocal (or off-diagonal with respect to the atomic sites)
contributions and, for the normal insulators studied in this work, these nonlocal contributions
will be substantially larger than the local
(or site-diagonal) ones. This is one of the most interesting aspects of the modern theory
of the orbital magnetization, which raises many new questions. Particularly, can these large nonlocal
contributions be measured or can they contribute to other properties?

  Another interesting issue is related to the first fundamental question -- the direction for the
improvement of SDFT. Will
this large and essentially nonlocal part of the orbital magnetization contribute to the exchange-correlation energy,
for instance -- in the framework of frequently discussed in this context current SDFT?\cite{VignaleRasolt,Ebert,ShiPRL}
Unfortunately, the explicit form of the exchange-correlation energy in terms of these orbitals currents is
largely unknown, and today
it is an open (but very interesting) question whether such theory can also
improve the description of local orbital magnetization, which is probed by many experiments.
As was pointed our in the Introduction, so far the dominant point of view was that the orbital magnetization
is a local quantity and the main processes, which are missing in practical DFT calculations and which are
responsible for the agreement with the experimental data can be also formulated in the local
(or site-diagonal) form.\cite{OPB,Norman,LDAU,Minar} In the light of this new funding, how general is this
conclusion and how important are the non-local processes, associated with the appreciable
$\textbf{k}$-dependence of the itinerant part of the orbital magnetization?

\textit{Acknowledgements}.
This work is partly supported by the grant of the Ministry of Education and Science of Russia N 14.A18.21.0889.


\begin{thebibliography}{99}
\bibitem{remark2}
The relativistic SO coupling is proportional to gradient of the scalar potential.\cite{White}
Therefore, in order to produce finite SO coupling, this potential should be position-dependent (or ``inhomogeneous'').
The external vector potential is also position-dependent, even for the uniform magnetic field.
\bibitem{remark1}
In the G$_{\boldsymbol{a}}$-A$_{\boldsymbol{b}}$-F$_{\boldsymbol{c}}$ ground state,
the twofold rotation around the orthorhombic $\boldsymbol{c}$ axis enters as it is,
while two other rotations around the orthorhombic $\boldsymbol{a}$ and $\boldsymbol{b}$
axes enter in the combination with the time inversion.
\end{thebibliography}
\end{document}